\documentclass[prd,showpacs,preprint]{revtex4}
\usepackage{amsfonts,amsmath,amssymb,revsymb}
\usepackage{hyperref}
\newcommand{\be}{\begin{equation}}
\newcommand{\ee}{\end{equation}}

\newcommand{\bea}{\begin{eqnarray}}
\newcommand{\eea}{\end{eqnarray}}

\newcommand{\bes}{\begin{subequations}}
\newcommand{\ees}{\end{subequations}}

\newcommand{\nn}{\nonumber} 

\newcommand{\ra}{\longrightarrow}
\newcommand{\cN}{{\cal N}}

\newcommand{\Tr}{\mbox{Tr}}
\newcommand{\tr}{\mbox{tr}}
\newcommand{\cZ}{{\cal Z}}
\newcommand{\ga}[1]{ A^{(#1)} }
\newcommand{\gau}[3]{ A^{(#1)}_{ \vec{#2}\vec{#3} } }
\begin{document}
\title{Quiver Chern-Simons theories and 3-algebra orbifolds}
\date{\today}
\author{Nakwoo Kim}
\affiliation{Department of Physics and Research Institute of Basic Science, \\
Kyung Hee University, Seoul 130-701, Korea}
\email{nkim@khu.ac.kr}

\begin{abstract}
We attempt to derive quiver Chern-Simons-matter theories
from the Bagger-Lambert theory with Nambu bracket, 
through an orbifold prescription which effectively induces
a dimensional reduction of the internal space for 3-algebra. 
We consider M2-branes on 
an ${\cal N}=4$ orbifold $\mathbb{C}^2/\mathbb{Z}_k
\times \mathbb{C}^2$, and compare the result with the so-called
dual Aharony-Bergman-Jafferis-Maldacena model, proposed recently 
by Hanany, Vegh, and Zaffaroni. Unlike the ${\cal N}=6$ example
$\mathbb{C}^4/\mathbb{Z}_k$, 
we find ambiguities in the matrix regularization. 
\end{abstract}
\maketitle

\section{Introduction}
The AdS/CFT correspondence \cite{Maldacena:1997re}
suggests that there exist a three 
dimensional superconformal field theory which is dual to M-theory
in $AdS_4\times S^7$ background. For a long time the gauge field
theory for multiple
membranes has remained a mystery, but recently a lot of 
progress has been made through the study of Chern-Simons-matter 
theory as a candidate description.

Bagger and Lambert \cite{Bagger:2006sk,Bagger:2007jr,Bagger:2007vi}, and also independently  Gustavsson \cite{Gustavsson:2007vu}, 
constructed
new maximally supersymmetric gauge theories in three 
dimensions.
These theories are based on a 3-algebra: unlike the familiar
Yang-Mills symmetry, 3-algebra naturally induces quartic 
Yukawa-type interaction, while scalar fields exhibit sixth-order
potential. The relevance of 3-algebra in membrane theory is
justified by the Myers effect \cite{Myers:1999ps}, 
which predicts multiple membranes in 
external field polarize into M5-branes.

When one combines 3-algebra with maximal supersymmetry, the
mathematical consistency condition turned out to be too
stringent. There exist only one positive-definite, finite
dimensional 3-algebra
\cite{Papadopoulos:2008sk,Gauntlett:2008uf}.
And then the entire Bagger-Lambert-Gustavsson (BLG) theory 
simplifies to Chern-Simons-matter theory with $SU(2)\times
SU(2)$ gauge symmetry, with levels $k$ and $-k$, and four 
bifundamental chiral multiplets with quartic superpotential.

On the other hand, Aharony, Bergman, Jafferis and Maldacena
(ABJM)
constructed new $\cN=6$ superconformal Chern-Simons-matter 
theories, which describe multiple membranes on orbifold 
$\mathbb{C}^4/\mathbb{Z}_k$ \cite{Aharony:2008ug}. 
Instead of relying on 
3-algebra as a new dynamical principle, they started with
 certain IIB brane configurations and utilized string duality
 relations and identified the gauge field theory living on 
 brane intersections. Unlike BLG theory, 
 the ABJM model can describe an arbitrary
 number of membranes: when there are $N$ M2-branes, the theory 
 has $U(N)\times U(N)$ gauge symmetry with level $k$ and $-k$.

It is then natural to ask how to generalize to 
different orbifolds. One can try to devise sophisticated 
brane configurations which would 
lead to less supersymmetric orbifolds \cite{Imamura:2008nn}.
Or one scans the set of quiver Chern-Simons-matter theories
to discover new examples of duality between gauge theory and
geometric singularities \cite{Hanany:2008fj}. More systematically
one should make use of brane tiling \cite{Hanany:2008cd}
or brane crystal \cite{Lee:2006hw,Imamura:2008qs} techniques.

In this article we take a different route. We start with 
the infinite dimensional 3-algebra, which is the Nambu bracket
of functions on $T^3$. The physical meaning of such theory
with an infinite number of fields is rather unclear, although
one might conceive it is suggestive of higher-dimensional M-theory 
branes \cite{Ho:2008nn,Jeon:2008bx,Park:2008qe,Jeon:2008zj}. In this paper we use the Nambu bracket 
theory as a technical tool and impose a 3-algebra version 
of orbifold truncation which effectively reduces $T^3
\to T^2$. This technique has been applied to $\mathbb{C}^4/\mathbb{Z}_k$ and correctly generated the large-$N$ 
limit of the ABJM model \cite{Kim:2008gn}. 

We study specifically a ${\cN=4}$ orbifold $\mathbb{C}^2/\mathbb{Z}_k\times \mathbb{C}^2$ in this article.
In fact a quiver gauge theory has been already proposed for this
example, and is dubbed a dual-ABJM model \cite{Hanany:2008fj}.
It again possesses $U(N)\times U(N)$ with levels $k,-k$, but with
a different matter content. One hypermultiplet is in 
bi-fundamental, and the other is in adjoint representation of $U(N)$.

Although the mesonic vacuum moduli
space is certainly $\mathbb{C}^2/\mathbb{Z}_k\times \mathbb{C}^2$, 
the correspondence is rather subtle. First of all, the
field theory itself has only ${\cN=3}$ supersymmetry, and the 
scalar potential fails to exhibit $SO(4)$ symmetry
\cite{RodriguezGomez:2009ae}.  
The superconformal index computation largely agrees with supergravity
result but not completely, although it is still open that the
twisted sector contribution might be responsible for 
the mismatch \cite{Choi:2008za}.

Using the Nambu bracket we will see that one
can reproduce the dual-ABJM model, but there is a caveat. 
For the adjoint matter one has to ignore terms which converge more
slowly than in the rest of the action, when taking the large-$N$ 
limit. Our analysis hopefully provides a different perspective for
the validity of dual-ABJM conjecture. 

In Sec.\ref{2} we review the BLG theory and the Nambu bracket
orbifold procedure. Sec.\ref{3} is the main part which studies
$\mathbb{C}^2/\mathbb{Z}_k\times \mathbb{C}^2$ orbifolds. We conclude
in Sec.\ref{4} with discussions.



\section{From 3-algebra to quiver gauge theory}
\label{2}
We start by briefly reviewing the Bagger-Lambert theory \cite{Bagger:2006sk,Bagger:2007jr,Bagger:2007vi} 
and the orbifold proposal made in Ref.~\cite{Kim:2008gn}. 
The $D=3,\, \cN=8$ superconformal field theory \cite{Bagger:2007vi} is based on a 3-algebra. 
For basis vectors $T^a$, the structure constants $f^{abcd}$ are defined through a 3-product as 
\be
[ T^a, T^b, T^c ] = f^{abc}_{\;\;\;\;\;\, d} T^d .
\ee

We are particularly interested in the Nambu bracket in this paper. The elements of 
3-algebra are scalar functions on a compact Riemannian 3-manifold $\Sigma$. With metric $g$
and coordinates $\sigma_i, \, i=1,2,3$, the Nambu bracket is 
\be
[X, Y, Z] = \frac{1}{\sqrt{g}}\epsilon^{ijk} \partial_i X \partial_j Y \partial_k Z . 
\ee

In this paper we choose $\Sigma=T^3$ with $g_{ij}=\delta_{ij}$, 
and set the period of $\sigma_i$ to be $2\pi$. 
The inner product of two elements in this 3-algebra is then naturally defined as
\be
(X, Y) \equiv \tr (X^* Y) = \frac{1}{(2\pi)^3} \int d^3 \sigma X^* Y .
\ee
If we choose the basis vectors $X_{\vec{M}}$ to be associated with the 
Fourier mode $e^{i\vec{M}\cdot\vec{\sigma}}$, the structure constants are easily 
computed 
\be
f_{\vec{N}_1, \vec{N}_2 , \vec{N}_3 , \vec{N}_4} = i \kappa ( \vec{N}_1 \times \vec{N}_2 \cdot \vec{N}_3 ) \,
\delta ( \vec{N}_1+ \vec{N}_2 + \vec{N}_3 + \vec{N}_4 ) ,
\label{cc}
\ee
where $\kappa$ is a tunable coupling constant. 

As usual with supersymmetric gauge theory, the Bagger-Lambert action is determined once we specify the matter content, 
the gauge transformation rules and the superpotential. There are four chiral multiplets, and 
the four complex scalar fields $Z^I, I=1,2,3,4$ couple to the gauge fields as follows
\be
D_\mu Z^I_a = \partial_\mu Z^I_a - \tilde{A}^{\;\;b}_{\mu\;\; a} Z^I_b
, \quad \tilde{A}^{\;\;b}_{\mu\;\; a} = f^{cdb}_{\;\;\;\;\;\; a} A_{\mu cd} . 
\ee
And the superpotential is given as 
\be
W =   2\,  \tr ( [Z^1 , Z^2 , Z^3 ] Z^4 ) . 
\label{sup}
\ee

Instead of working with the entire Lagrangian, we will just consider how the truncation and 
regularization process affect the bosonic sector, i.e. 
scalar fields and gauge field. We will
also analyze their 
interactions through the gauge coupling and superpotential. Let us just quote here the 
Chern-Simons kinetic term 
\be
\frac{1}{2} \varepsilon^{\mu\nu\lambda}
\left[ f^{abcd} A_{\mu ab} \partial_\nu A_{\lambda cd} +
\frac{2}{3} f^{cda}_{\;\;\;\;\;\; g} f^{efgb} A_{\mu ab}
A_{\nu cd} A_{\lambda e f}
\right] , 
\label{cs}
\ee
which will be essential in identifying the gauge group structure.

It is $\cN=6$ orbifold $\mathbb{C}^4/\mathbb{Z}_k$ which was considered in Ref.~\cite{Kim:2008gn}. 
The spacetime symmetry is combined with translation along $\sigma_3$, and in the end 
one keeps only the modes with 
momentum number $\alpha_I=(1,1,-1,-1)$
along $\sigma_3$ for $Z^I$. It can be facilitated also
if we formally demand invariance for all $k$, i.e.
\be
{Z}^I (\sigma_1 , \sigma_2 , \sigma_3 ) = e^{2\pi i\alpha_I/k} {Z}^I (\sigma_1 , \sigma_2 , \sigma_3 -2\pi/k ) \quad \mbox{for all $k>0$}.
\label{1pr}
\ee
Note that the internal space $T^3$ is now effectively reduced to $T^2$. It is convenient to denote for 
instance as $Z^1_{\vec{m}}\equiv Z^1_{(\vec{m},1)}$, 
$A^\mu_{\vec{m}\vec{n}}\equiv A^\mu_{(\vec{m},1),(\vec{n},-1)}$, 
where $\vec{m},\vec{n}\in \mathbb{Z}^2$.

Now the action should be expressed as a 
(multiple) summation over these $\mathbb{Z}^2$ indices, which can be regularized in terms of 
large matrices. For $N\times N$ matrices, one first introduces clock and shift matrices
($\xi = e^{2\pi i/N}$)
\be
UV = \xi VU , \quad
U^N = V^N = 1 .
\ee
Basis vectors are defined and their multiplication is easily computed as follows
\bea
J^{\vec{n}} = \xi^{-\frac{n_1 n_2}{2}} U^{n_1} V^{n_2} , 
\quad \quad 
J^{\vec{m}} J^{\vec{n}} = \xi^{\vec{m}\times \vec{n}/2} J^{\vec{m}+\vec{n}} .
\eea

In order to regularize the infinite dimensional 
3-algebra action, we construct the matrix fields as 
\be
Z^I = \frac{1}{\sqrt{N}} \sum_{\vec{m}} Z^I_{\vec{m}} {J^{\vec{m}}}, 
\quad \quad 
A^{\pm}_{\mu} = \frac{4\pi i}{kN} \sum_{\vec{p},\vec{q}} 
\xi^{\pm \vec{p}\times\vec{q}/2} A_{\mu\vec{p}\vec{q}} 
J^{\vec{p}+\vec{q}}. 
\label{param}
\ee
Then one can verify \cite{Kim:2008gn} that the large-$N$ limit of
ABJM model with $U(N)\times U(N)$ gauge group and four bifundamental 
chiral multiplets is equivalent to the truncated 3-algebra theory 
through Eq.~(\ref{param}). The original coupling constant $\kappa$
in that case is given in terms of gauge theory parameters as  
\be
\kappa = - \frac{4\pi^2}{kN^2} .
\label{coup}
\ee

\section{$\cN=4$ orbifold $\mathbb{C}^2/\mathbb{Z}_k \times \mathbb{C}^2$}
\label{3}
Let us now move to our main interest, $\mathbb{C}^2/\mathbb{Z}_k \times \mathbb{C}^2$. 
The discrete symmetry action is defined on $\mathbb{C}^4$ as 
\be
(Z^1, Z^2, Z^3, Z^4) \ra ( \omega Z^1, \omega^{-1} Z^2 , Z^3 , Z^4 ),
\ee
where $\omega=e^{2\pi i /k}$. 
In this 
background the M2-brane theory should preserve $\cN=4$, i.e. half of the maximal supersymmetry. 
As before we start with the BL theory with Nambu bracket on $T^3$, and restrict the momentum modes.
The above transformation will be 
combined with $\mathbb{Z}_k$ translation in the internal space. With $\alpha_I=(1,-1,0,0)$, we demand
\be
Z^I( \sigma_1 , \sigma_2 , \sigma_3 ) = e^{2\pi\alpha_I/k} Z^I ( \sigma_1 , \sigma_2 , \sigma_3 
- 2\pi/k )
, \quad \mbox{for all $k>0$} .
\ee

We thus fix the 
momentum modes of $Z^I$ along $\sigma_3$ to be $\alpha_I$, and write as 
\be
Z^I (x^\mu ; \sigma_i)  = \sum_{\vec{m}} \cZ^I_{\vec{m}} (x^\mu)
 e^{i\vec{m}\cdot\vec{\sigma}} e^{i\alpha_I \sigma_3} . 
\ee
where $\vec{\sigma}=(\sigma_1,\sigma_2)$. 
One can again construct four matrix fields using $\cZ^I_{\vec{m}}$ as the expansion 
coefficients for the basis vector $J^{\vec{m}}/\sqrt{N}$. Obviously $Z^3,Z^4$ should
take a different representation than $Z^1,Z^2$, so we introduce a new symbol and write
$\Phi_1 = \cZ^3_{\vec{m}} J^{\vec{m}}/\sqrt{N},\, \Phi_2 = \cZ^4_{\vec{m}} J^{\vec{m}}/\sqrt{N}$.

For the consistency we should keep the same type of modes 
for the gauge field 
$A^\mu_{\vec{M}\vec{N}}$. 
Since the covariant derivative should not violate the orbifold projection, we have two 
surviving sets: one is $(M_3,N_3)=(1,-1)$ or $(-1,1)$, and the other possibility is $(0,0)$. 
We introduce double-index gauge fields as 
\be
A^{(1)}_{\vec{m}\vec{n}} = A_{(\vec{m},1)(\vec{n},-1)}, \quad A^{(0)}_{\vec{m}\vec{n}}
= A_{(\vec{m},0)(\vec{n},0)} . 
\ee
One can show $A^{(1)*}_{\vec{m}\vec{n}} = 
A^{(1)}_{-\vec{m},-\vec{n}}$ and $A^{(0)*}_{\vec{m}\vec{n}} = +
A^{(0)}_{-\vec{m},-\vec{n}}$, from antisymmetry and reality condition of $A_{\vec{M}\vec{N}}$.

In order to identify the gauge symmetry of the truncated and regularized 3-algebra action, 
let us first consider the Chern-Simons type kinetic term, 
given above in Eq.~(\ref{cs}). One can easily verify
\bea 
\frac{1}{2} \varepsilon^{\mu\nu\lambda}
f^{abcd} A_{\mu ab}  \partial_{\nu} A_{\lambda cd} 
&=& -4i\kappa \sum_{\vec{m}+\vec{n}+\vec{p}+\vec{q}=0}
\left[ 
(\vec{p}\times\vec{q})
A^{(1)}_{\vec{m}\vec{n}} \mbox{d} A^{(0)}_{\vec{p}\vec{q}}
\right.
\nn\\
&&
\left.
+(\vec{m}\times \vec{n}+\vec{p}\times\vec{q})
A^{(1)}_{\vec{m}\vec{n}} \mbox{d} A^{(1)}_{\vec{p}\vec{q}}
\right] . 
\label{csk}
\eea

We again want to construct matrix fields using the 3-algebra 
expansion coefficients $A^{(I)}_{\vec{p}\vec{q}}$. 
As a start, let us define matrix gauge fields in the same way as we did for $\mathbb{C}^4/\mathbb{Z}_k$.
\bea
A^{\pm(I)}_{\mu} = \frac{4\pi i}{kN} 
 \sum \xi^{\pm\vec{p}\times\vec{q}/2}
A^{(I)}_{\mu\vec{p}\vec{q}} J^{\vec{p}+\vec{q}} 
, \quad I=0,1. 
\eea
In fact one can show $A^{+(0)} = - A^{-(0)}$, using $A^{(0)}_{\vec{m}\vec{n}} = -
A^{(0)}_{\vec{n}\vec{m}}$. Now we consider some products of matrix fields and get 
the following results:
\bea
\Tr ( \ga{1}_+ \mbox{d} \ga{1}_+ - \ga{1}_- \mbox{d} \ga{1}_- )&=& -\frac{32\pi^3 i}{k^2N^2}
\sum_{\vec{m}+\vec{n}+\vec{p}+\vec{q}=0} 
( \vec{m}\times \vec{n}+\vec{p}\times \vec{q} ) \gau{1}{m}{n} \,\mbox{d} \gau{1}{p}{q}
\\
\Tr \,(A^{(1)}_+ +A^{(1)}_-) \,\mbox{d} A^{(0)}_+ &=& 
-\frac{32\pi^3 i}{k^2N^2} 
\sum_{\vec{m}+\vec{n}+\vec{p}+\vec{q}=0} 
 (\vec{p}\times \vec{q}) \, 
\gau{1}{m}{n} \,\mbox{d} \gau{0}{p}{q}
\eea
Then it is easy to see that, if we define new matrix fields as 
\be
\tilde{A}^{\pm} = A^{\pm(1)} + \tfrac{1}{2} A^{\pm(0)} , 
\ee
Eq.~(\ref{csk}) is reproduced by 
\be
\tfrac{k}{4\pi} \Tr ( \tilde{A}_+ \mbox{d} \tilde{A}_+ - \tilde{A}_- \mbox{d} \tilde{A}_- )
\label{cs_quad}
\ee 
in the large $N$ limit, if we again demand Eq.~(\ref{coup}). 

Compared to the $\cN=6$ case, the truncation seems to give 
us one more
gauge field $A^{(0)}$, but rather surprisingly the study of 
gauge kinetic term suggests we still have $U(N)\times U(N)$ gauge
groups with Chern-Simons level $k$ and $-k$ respectively. 
Of course we need to check whether this simplification persists
with the gauge field cubic interaction term. The second term 
in Eq.~(\ref{cs}) is reduced by the truncation into  
\be
4\kappa^2 \sum_{\vec{m}+\vec{n}+\vec{p}+\vec{q}+\vec{r}+\vec{s}=0}
\left[(\vec{m}+\vec{n})\times(\vec{p}+\vec{q})\right]
(\vec{r}\times\vec{s})
\gau{1}{m}{n} \gau{1}{p}{q} \left( \gau{1}{r}{s} +
\gau{0}{r}{s}/2 \right) . 
\label{3alg_cubic}
\ee

Let us then consider the cubic term which would be 
consistent with Eq.~(\ref{cs_quad}) on the matrix side:
\bea
\Tr ( \tilde{A}^3_+ - \tilde{A}^3_- )
&=& \Tr \left( A^{(1)3}_+ - A^{(1)3}_- \right)
+\tfrac{3}{2} \Tr \, \ga{0}_+ \left( A^{(1)2}_+ + A^{(1)2}_- \right)
\nn\\
&&
+\tfrac{3}{4} \Tr A^{(0)2}_+ \left( \ga{1}_+ - \ga{1}_- \right)
+ \tfrac{1}{4} \Tr A^{(0)3}_+ . 
\eea
One can easily see that, in the large-$N$ limit the first two 
terms are ${\cal O}(1/N)$, while the last two terms are
${\cal O}(1/N^3)$ and thus negligible in the above expression. 
Now it is rather tedious but straightforward to show that 
$\frac{2i}{3}\frac{k}{4\pi}\Tr ( \tilde{A}^3_+ - \tilde{A}^3_- )$ 
in the large-$N$ limit is equivalent to  
Eq.~(\ref{3alg_cubic}), if we accept Eq.~(\ref{coup}).
One can thus establish that for our $\cN=4$ orbifold the matrix
regularization reduces the Chern-Simons kinetic terms into 
\be
\tfrac{k}{4\pi} \Tr ( \tilde{A}_+ \mbox{d} \tilde{A}_+ 
+\tfrac{2i}{3} \tilde{A}_+^3 
- \tilde{A}_- \mbox{d} \tilde{A}_- 
-\tfrac{2i}{3} \tilde{A}_-^3 ) . 
\ee

We now turn to check if the matter fields have consistent couplings
with $\tilde{A}_{\pm}$. For $Z^1$($Z^2$ is simply 
the complex conjugate), we have
\be
(D_\mu {\cal Z}^1)_{\vec{n}}
= \partial_\mu {\cal Z}^1_{\vec{n}}
+  i \kappa 
\sum_{\vec{p}+\vec{q}+\vec{m}=\vec{n}}
\left[
2\left(
\vec{p}\times \vec{q} + (\vec{p}+\vec{q})\times \vec{m}
\right) A^{(1)}_{\mu\vec{p}\vec{q}} 
+(\vec{p}\times \vec{q}) A^{(0)}_{\mu\vec{p}\vec{q}} 
\right]
{\cal Z}^1_{\vec{m}} . 
\label{scalar_coupling}
\ee
Let us define a scalar field on the matrix side again as 
${\cal Z}^1 = 
\tfrac{1}{\sqrt{N}} \sum_{\vec{m}} {\cal Z}^1_{\vec{m}} J^{\vec{m}}$,
with a slight abuse of notation.
It is then straightforward to verify that, the matrix field representation 
\be
D_\mu{\cal Z}^1 = 
\partial_\mu {\cal Z}^1
+  i 
( \tilde{A}_{+\mu} {\cal Z}^1 - {\cal Z}^1 \tilde{A}_{-\mu} )  
\ee
in the large-$N$ limit approaches 
Eq.~(\ref{scalar_coupling}), with the same identification 
Eq.~(\ref{coup}). This obviously implies ${\cal Z}^1$
is in $({\bf N}, {\bf \bar{N}})$ representation of 
$U(N)\times U(N)$, while ${\cal Z}^2$ is $({\bf \bar{N}}, {\bf {N}})$.

The remaining scalar fields $\Phi^1,\Phi^2$ on the other hand
couple to $A^{(1)}$ only, and the 3-algebra side computation
gives 
\be
(D_{\mu} \Phi)_{\vec{n}} = 
\partial_{\mu} \Phi_{\vec{n}} + 2 i \kappa 
\sum_{\vec{p}+\vec{q}+\vec{m}=\vec{n}}
\left[ (\vec{p}+\vec{q})\times \vec{m} \right]
A^{(1)}_{\mu\vec{p}\vec{q}} \Phi^{}_{\vec{m}} . 
\label{adj_cov}
\ee
Our objective here is to identify the
matrix Chern-Simons-matter theory which approaches
the truncated 3-algebra theory in the large-$N$ limit. The correction 
terms, or the error, have been kept small
 by ${\cal O}(1/N^2)$, compared to the leading order terms. 
 It is natural 
to require the same with $\Phi$ fields, but 
then the covariant derivative has to be 
\be
D_\mu \Phi
= \partial_\mu \Phi + i \left[
 \frac{\tilde{A}_{+\mu} + \tilde{A}_{-\mu}}{2} \Phi -
  \Phi \frac{\tilde{A}_{+\mu} + \tilde{A}_{-\mu}}{2} \right] .
\label{phi_cov}
\ee
It is only with the above expression that we can cancel the contributions from $A^{(0)}$ in 
$\tilde{A}_{\pm}$, and correctly reproduce Eq.~(\ref{adj_cov})
up to ${\cal O}(1/N)$. However it is obvious that the gauge 
coupling of $\Phi$ in 
Eq.~(\ref{phi_cov}) does not exhibit a well-defined transformation
rule. In other words, Eq.~(\ref{adj_cov}) is not compatible with 
$U(N)\times U(N)$ gauge symmetry represented by $\tilde{A}_{\pm}$.

If we insist on consistent gauge symmetry, we may instead choose
\be
D_\mu \Phi
= \partial_\mu \Phi + i \left(
 \tilde{A}_{+\mu} \Phi -
  \Phi \tilde{A}_{+\mu}  \right) .
\label{avz}
\ee
Then $\Phi$'s are in the adjoin representation of $\tilde{A}_+$, 
but now the correction terms are order ${\cal O}(1/N)$! Note that
one can substitute $\tilde{A}_+$ with $\tilde{A}_-$ in Eq.(\ref{avz}),
and again the terms involving $A^{(0)}_{\vec{m}\vec{n}}$ are smaller
by ${\cal O}(1/N)$, compared to $A^{(1)}$ terms. One should note
the difference with ${\cal Z}$ fields, where the coupling is
$(\tilde{A}_+ {\cal Z} - {\cal Z} \tilde{A}_-)$ and the correction 
terms are ${\cal O}(1/N^2)$.

Finally let us consider the superpotential. Using the same 
matrix parametrization, it is 
\be
W = \frac{4\pi}{k} \Tr ( {\cal Z}^1 {\cal Z}^2 [ \Phi^1 , 
\Phi^2 ] )
\ee
which is consistent with the gauge symmetry and approaches 
Eq.~(\ref{sup})
in the large-$N$ limit. Again, the correction terms are
relatively ${\cal O}(1/N)$. 

With the above prescription
 our orbifolded action becomes equivalent to the
quiver gauge theory for membranes on 
$\mathbb{C}^2/\mathbb{Z}_k \times \mathbb{C}^2$ 
proposed in Ref.\cite{Hanany:2008fj}.

\section{Discussion}
\label{4}
In this article we continued with our proposal 
\cite{Kim:2008gn} for 3-algebra theory and applied it
to a ${\cN=4}$ orbifold. The Nambu-bracket BLG theory has an 
infinite number of fields, and its physical relevance is unclear
at the quantum level. But it certainly enjoys maximal 
supersymmetry, so one might fathom it somehow encompasses the 
multi-membrane theory in flat background. Indeed, it can be shown 
that when one takes a version of orbifold procedure 
$\mathbb{C}^4/\mathbb{Z}_k$ the 3-algebra
theory becomes equivalent to large-$N$ limit of the ABJM model. 

For the $\mathbb{C}^2/\mathbb{Z}_k \times \mathbb{C}^2$ orbifold
we have witnessed a problem with matrix regularization. If 
we express the 3-product using the Poisson bracket in the most natural 
way, the matrix regularization does not lead to a consistent
Yang-Mills invariant theory. One needs to sacrifice the consistency 
of large-$N$ approximation and allow for ${\cal O}(1/N)$ discrepancy
which was not present with $\mathbb{C}^4/\mathbb{Z}_k$.

Our approach might explain the puzzle of supersymmetry enumeration 
with dual ABJM model. Although it is presented as a
dual to ${\cN=4}$ orbifold, the field theory model has
 only ${\cN=3}$. In our 
analysis, the reduced 3-algebra description, written for instance
using Eqs.~(\ref{csk}),(\ref{3alg_cubic}),(\ref{scalar_coupling}),
and (\ref{adj_cov}) must be ${\cal N}=4$. But the 3-algebra side
action 
corresponds to the large-$N$ limit of dual ABJM model, at best. 
It should be the mismatch
of orders of magnitude for subleading terms which break the 
supersymmetry to ${\cN=3}$.

Finally, one might ask why we have an ambiguity in identifying the
gauge symmetry. After all, the BLG theory is invariant under
3-algebra transformation. We can think of two technical points.
First, the 3-algebra
symmetry is presented only with infinitesimal transformation
in Ref.~\cite{Bagger:2006sk,Bagger:2007jr,Bagger:2007vi}: 
unlike Yang-Mills symmetry, there is no analogue of multiplication 
by generic unitary matrix or finite transformation {\it per se}. 
This problem is especially acute when
we see Eq.~(\ref{phi_cov}), which would suffer from non-commutativity
of unitary transformations for $\tilde{A}_+$ and $\tilde{A}_-$. 
Second, the supersymmetry and
3-algebra transformation rules of BLG theory are given for 
$\tilde{A}^{\;\;b}_{\mu\;\; a} = f^{cdb}_{\;\;\;\;\;\; a} A_{\mu cd}$, 
while the action itself is given in terms of $\tilde{A}^{\;\;b}_{\mu\;\; a}$. For $SO(4)$ one can immediately invert and express the
transformation rules for $\tilde{A}^{\;\;b}_{\mu\;\; a}$, but with 
Eq.~(\ref{cc}) the inversion is far from clear.

\acknowledgments
The research of N. Kim is supported by the Science Research Center Program of the Korea Science and Engineering Foundation through the Center for Quantum Spacetime (CQUeST) of Sogang University with grant number R11-2005-021, and also partly by the Korea Research Foundation Grant No. KRF-2007-331-C00072.

\bibliography{prd_db}{}
\end{document}